\documentclass[%
 reprint,
superscriptaddress,
 amsmath,amssymb,
 aps,prl,
]{revtex4-1}

\usepackage{todonotes}
\usepackage{graphicx}% Include figure files
\usepackage{dcolumn}% Align table columns on decimal point
\usepackage{bm}% bold math
\usepackage[normalem]{ulem}
\usepackage{microtype}
\usepackage{xcolor}
\definecolor{DragonGreen}{RGB}{0,126,48}
\definecolor{Brownie}{RGB}{97,16,9}

\usepackage[paperwidth=210mm,
            paperheight=297mm,
            left=20mm,
            top=10mm,
            textwidth=170mm,
            marginparsep=3mm,
            marginparwidth=30mm,
            textheight=730pt,
            footskip=50pt]
           {geometry}

\newcommand{\gsl}{\ensuremath{\gamma_{\text{sl}}}}
\newcommand{\ns}{\ensuremath{n_{\text{s}}}} % number of solid-like atoms
\newcommand{\nl}{\ensuremath{n_{\text{l}}}}
\newcommand{\ms}{\ensuremath{\mu_{\text{s}}}} % chemical potential
\newcommand{\ml}{\ensuremath{\mu_{\text{l}}}}
\newcommand{\msl}{\ensuremath{\mu_{\text{sl}}}} %
\newcommand{\phis}{\ensuremath{\phi_{\text{s}}}} % order p
\newcommand{\phil}{\ensuremath{\phi_{\text{l}}}}
\newcommand{\tm}{\ensuremath{T_{\text{m}}}}

\begin{document}

\preprint{APS/123-QED}

\title{The solid-liquid interfacial free energy out of equilibrium}

\author{Bingqing Cheng}
 \affiliation{Laboratory of Computational Science and Modeling, Institute of Materials, {\'E}cole Polytechnique F{\'e}d{\'e}rale de Lausanne, 1015 Lausanne, Switzerland}%

\author{Gareth A. Tribello}
\affiliation{Atomistic Simulation Centre, School of Mathematics and Physics, Queen's 
University Belfast, Belfast, BT7 1NN
}%

\author{Michele Ceriotti}
\email{michele.ceriotti@epfl.ch}
\affiliation{Laboratory of Computational Science and Modeling, Institute of Materials, {\'E}cole Polytechnique F{\'e}d{\'e}rale de Lausanne, 1015 Lausanne, Switzerland}%

\date{\today}

\begin{abstract}

%{\bf ALTERNATIVE ABSTRACT } 
The properties of the interface between solid and melt are key to
solidification and melting, as the interfacial free 
energy introduces a kinetic barrier to phase transitions. 
This makes solidification happen below the melting 
temperature, in out-of-equilibrium conditions at
which the interfacial free energy is ill-defined. 
Here we draw a connection between the atomistic 
description of a diffuse solid-liquid interface
and its thermodynamic characterization. This 
framework resolves the ambiguities in defining 
the solid-liquid interfacial free energy 
above and below the melting temperature. 
In addition, we introduce a simulation 
protocol that allows solid-liquid 
interfaces to be reversibly created and 
destroyed at conditions relevant 
for experiments. We directly evaluate the value 
of the interfacial free energy away from the 
melting point for a simple but realistic atomic potential, 
and find a more complex temperature dependence than
the constant positive slope that has been generally assumed 
based on phenomenological considerations and that has been used to
interpret experiments. This methodology could be easily 
extended to the study of other phase transitions,
from condensation to precipitation.
Our analysis can help reconcile the textbook picture 
of classical nucleation theory with the growing body 
of atomistic studies and mesoscale models of solidification.  

%The properties of the solid-liquid interface underpin
%solidification and melting, but are exceedingly hard to
%investigate because these processes typically occur in out-of-equilibrium
%conditions. Here we introduce an atomistic simulation protocol 
%and a consistent thermodynamic
%framework that allow solid-liquid interfaces to
%be studied at conditions relevant for experiments
%and mesoscale models of solidification. 
%We use this technique to directly evaluate the value and the temperature dependence
%of the solid-liquid interface free energy for a simple but
%realistic Lennard-Jones system, and show that
%the positive temperature dependence that is generally assumed based on
%phenomenological considerations is far from universal. 
%Our framework can help reconcile
%the textbook picture of classical nucleation theory
%with an atomistic understanding of the structure and
%fluctuations of the interface between the two phases.
%This approach can be easily extended for other
%types of kinetically-hindered phase transitions.

\end{abstract}

\pacs{Valid PACS appear here}% PACS, the Physics and Astronomy
\maketitle

%\tableofcontents

Solidification underlies many natural phenomena 
such as the freezing of water in clouds and the
formation of igneous rock. 
It is also crucial for various
critical technologies including commercial casting, 
soldering and additive 
manufacturing~\cite{dantzig2009solidification,li+14ncomm,murphy20143d}.
Computational and experimental studies of these
phenomena are complicated by the fact that
homogeneous nucleation of solids often occurs at 
temperatures well below the melting point $\tm$~\cite{wu2004nucleation}.
The free energy change associated with the formation of a
solid nucleus containing $\ns$ particles is usually
written as
\begin{equation}
G(\ns) = \msl\ns + \gsl A(\ns).
\label{eq:G-ns}
\end{equation}
In this expression the first bulk term stems from
the difference in chemical potential between the solid
and the liquid $\msl=\ms-\ml$, which is negative 
below $\tm$. The second term describes 
the penalty associated with the interface 
between the two phases, and introduces
 a kinetic barrier to nucleation.
This surface term is the product of 
interfacial free energy $\gsl$, and the extensive surface
area $A(\ns)$. 
However, due to the diffuse nature 
of the interface, there is a degree of ambiguity 
in the location and area of the dividing surface between phases.
In classical nucleation theory (CNT),
an infinitesimally-thin dividing surface divides
the solid nucleus from the surrounding liquid. 
These two phases are usually taken to have their bulk densities so the surface area of the solid nucleus can be  calculated using
$A(\ns)=\sigma \ns^{2/3}$,
where $\sigma$ is a constant that depends on the shape, e.g.
$\sigma=(36\pi)^{1/3}v_\text{s}^{2/3}$ for a spherical nucleus with bulk solid molar volume $v_\text{s}$.
Under these assumptions, the rate of nucleation can be estimated by calculating the free-energy barrier to nucleation using $G^\star=\frac{4}{27} \sigma^3 \msl^{-2} \gsl^3$.

The experimental determination of nucleation rates
is challenging due to the non-equilibrium 
conditions, and is often 
affected by considerable uncertainties~\cite{hoyt2003atomistic,wu2004nucleation}. 
These difficulties have triggered the development of numerous 
computational methods for evaluating the free energy change during nucleation.
Inducing the formation of a solid nucleus from 
the melt in an atomistic simulation~\cite{ten1996numerical,ten1999homogeneous,trudu2006freezing}
allows one to directly evaluate $G(\ns)$ at any given temperature, without explicitly partitioning the free energy as in Eq.~\eqref{eq:G-ns}.
However, this 
approach is restricted by the system size, 
and usually only works at unphysical 
undercoolings, where the critical nucleus 
consists of a few hundred atoms.  
Extrapolating results 
for $G^\star$ obtained from these simulations to 
the mesoscale critical nuclei formed at experimental 
conditions inevitably requires invoking
models such as CNT, which involves quantities that are ill-defined
for nanosized nuclei~\cite{lechner2011role,prestipino2012systematic}. 
In addition, because 
the nuclei observed in these simulations 
are tiny, defining their shape and surface area
is problematic, which makes it almost impossible to 
infer the anisotropy of $\gsl$. 
For these reasons, an alternative strategy 
in which $\gsl$ is evaluated for 
an idealized planar interface, which resembles
the surface of a mesoscopic nucleus, is often adopted.
In the planar limit the value of the surface area $A(\ns)$ is
unequivocal.  In addition,
different crystallographic orientations
can be treated separately, so $\gsl$ and its anisotropy 
can be defined more rigorously. 
Techniques of this type include the 
capillary fluctuation method, which relies on 
an analysis of the height profile
for the atomically-rough solid-liquid 
interfaces~\cite{hoyt2001method,becker2009atomistic} and
the cleaving method, which computes the reversible work 
required to cleave the bulk solid and liquid and to join 
the pieces to generate an interface ~\cite{broughton1986molecular,davidchack2000direct,davidchack2003direct}.
These two methods only allow one to examine how the system behaves 
at the melting temperature, however.

To investigate the solid-liquid interface at
conditions that mirror the ones in experiments, 
a method that works
for planar interfaces
and that operates under 
realistic levels of undercooling is required.
In this paper, we derive just such a method 
by extending
an approach that uses a history-dependent bias
to reversibly create and dissolve a solid-liquid 
interface at equilibrium~\cite{angi+10prb,angi11jpcm}
so that it also works at  $T\ne \tm$.  We then use this method to 
compute the temperature dependence of $\gsl$
for a realistic interatomic potential. 

In order to relate atomic-scale descriptions 
with mesoscopic models and experiments, it is 
necessary to introduce a consistent framework
for quantifying the number of solid atoms $\ns$.
At the atomistic level, several families of 
structural fingerprints $\phi(i)$ 
can be used to determine whether the environment around atom $i$ is
solid-like or liquid-like \cite{stei+83prb}.
In our simulations, for example, we used a modified 
version of the cubic harmonic
order parameter $\kappa(i)$ introduced in 
Ref.~\cite{angi+10prb}, which, as described in the Supplemental Material at [URL will be inserted by publisher]~\cite{SI}, we further refined so as
to better discriminate between the liquid, the solid, 
and the various different crystal orientations  
by applying a switching function $S$
and defining $\phi(i)=S(\kappa(i))$.
Instead of defining $\ns$ by introducing 
an ad-hoc threshold on $\phi$, which is quite common
in simulations of this kind, we adopted an alternative approach, 
that is closely related to the classical 
definition of a Gibbs dividing surface based on a thermodynamic variable such as the volume $V$.
For each microstate, we construct a global collective variable (CV)
$\Phi=\sum_i \phi(i)$ by summing over all the atoms. 
Taking $\phis$ and $\phil$ to be the reference values of
the order parameter averaged over atoms in the bulk solid and the bulk 
liquid at the same thermodynamic conditions,
one can then define the number of atoms of solid as 
\begin{equation}
 \ns(\Phi) = \left(\Phi -N \phil\right)/
 \left(\phis-\phil\right). \label{eqn:nsl-phi}
\end{equation}
This is equivalent to choosing a reference state
that satisfies $\Phi = \phis \ns+ \phil \nl$, 
subject to the conservation of the total number of atoms in the 
system $\ns+\nl = N$. In other words,
zero surface excess for the extensive quantity $\Phi$ 
has been assumed.
Within this approach, the choice of the order parameter implicitly 
determines the position of the solid-liquid dividing surface while
automatically averaging over roughness and thermal fluctuations, which is essential for determining the macroscopic value of  
$\gsl$~\cite{prestipino2012systematic}.

We simulated the solid-liquid interface for 
a simple but realistic Lennard-Jones system~\cite{davidchack2003direct,angi+10prb,benjamin2014crystal}.
To accelerate the sampling so as to obtain reversible 
formation of a solid-liquid interface in a viable 
amount of simulation time,
we performed well-tempered metadynamics 
simulations with adaptive Gaussians~\cite{barducci2008well,branduardi2012metadynamics}.
using $\Phi=\sum_i S(\kappa(i))$ (see above) as the CV.
A simulation supercell with dimensions commensurate to the 
equilibrated lattice at the temperature $T$, was employed.  This cell was
elongated along the $x$ axis which was chosen to be parallel to the crystallographic orientation of the solid-liquid 
interface.
Simulations were run at 
constant temperature and pressure, with
only the $x$ direction left free to fluctuate.
Under these conditions, the planar interfaces, whenever present,
are always perpendicular to the chosen lattice direction and always have
a surface area 
$A=2\Delta y \Delta z$
equal to twice the supercell 
cross section.
The sampling has to be restricted to the relevant regions of phase space:
the melt, the coexistence state with correctly-oriented solid-liquid interfaces, and the defect-free fcc crystal.
Choosing an appropriate CV is critical for achieving this end and, in addition,
several judiciously tuned restraints are required to prevent the formation of
grain boundaries and twin defects. 
The presence of these defects was already
observed in biased simulations at $\tm$~\cite{angi11jpcm} and they become 
very likely at undercooled conditions. 
Fast implementation of this complex 
simulation setup was made possible by the flexibility of the PLUMED code~\cite{tribello2014plumed} in combination with LAMMPS~\cite{plim95jcp}.
See the Supplemental Material for sample input files and simulation details~\cite{SI}.

\begin{figure}
\centering
\includegraphics[width=0.95\columnwidth]{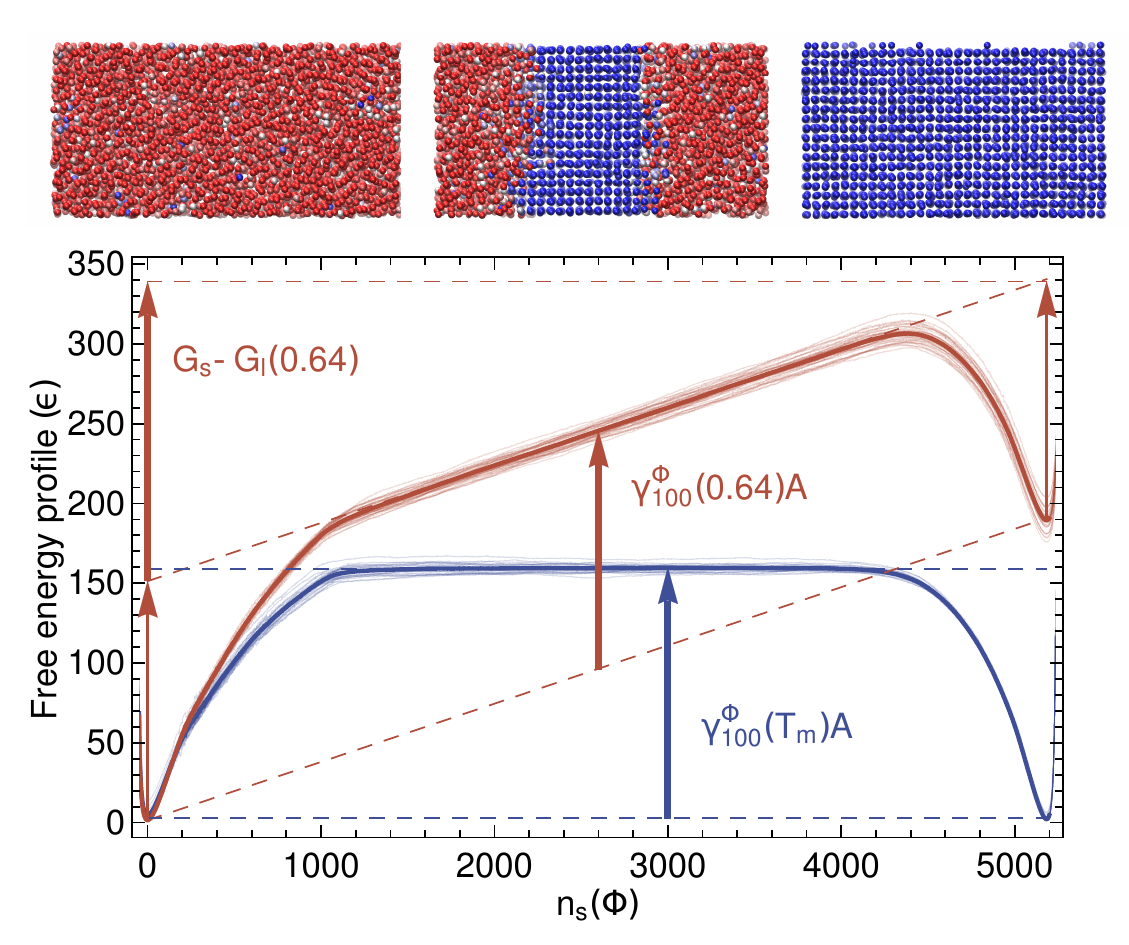}
    \caption{
    The free energy profile averaged from 36 independent metadynamics simulation runs, and shown 
    as a function of $\ns(\Phi)$, for an interface 
    perpendicular to $\left<100\right>$ and a simulation box    composed of
    $16\times 9\times 9$ \emph{fcc} unit cells.
    Representative atomic configurations for different values of 
    $\ns(\Phi)$ are shown, with atoms colored according to the local 
    order parameter $\phi$.
    The two sets of curves correspond to the free energy profiles
    at the melting temperature ($\gsl(T_m)=0.371(2)$) and 
    at a moderate degree of over-heating ($\gsl(0.64)=0.365(3)$).  All
quantities are expressed in Lennard-Jones units.
    \label{fig:fes1d}
}
\end{figure}
 
Figure~\ref{fig:fes1d} shows the 
free energy $G$ as a function of $\ns(\Phi)$ at two different temperatures.  These curves were
reconstructed from the metadynamics simulations using on-the-fly
re-weighting \cite{torr-vall99jcp,tiwa-parr14jpcb}. 
At $T=\tm$, $G(\ns(\Phi))$ displays
a broad horizontal plateau that corresponds to the progressive movement of the
solid-liquid interface.
As discussed in Ref.~\cite{angi+10prb}, $\gsl(\tm)A$ can be unambiguously
taken as the difference between the free energy of the plateau and the 
free energy at the bottom of the 
solid and liquid minima.
The bias potential allows the interface 
to form even when $\msl$ is non-zero, at temperatures
away from $\tm$.
Figure~\ref{fig:fes1d} thus also
reports the free energy profile computed for $T>T_m$. 
The free energies of the minima corresponding to the bulk solid and liquid 
phases now take different values. The slope of the line that joins these 
minima is equal to $\msl(T)$. The region that was a plateau at $T=T_m$
has become an oblique line, with a slope that is also equal to $\msl$ as
moving the dividing surface now has an energetic cost that is associated with 
forcing atoms to undergo the phase transition~\cite{pedersen2013computing}.
The vertical red arrow in Figure \ref{fig:fes1d}
indicates the free energy difference between the simulated
system with a solid-liquid interface and
a reference state composed of $\ns(\Phi)$ atoms 
of bulk solid and $N-\ns(\Phi)$ atoms 
of bulk liquid.  This quantity is equal to  $\gsl(T)A$.

At this point it is important to note that at $T\ne\tm$ 
a change in the definition of the order parameter $\phi$ 
affects how the free energy of a given state is partitioned
between the bulk and surface terms. We will demonstrate later
that the value we obtain for $\gsl(T)$ thus depends on $\phi$. 
As discussed in the Supplemental Material~\cite{SI}, this does not change the free energy 
barrier for the formation of a critical nucleus, provided that 
a consistent framework is used to define
the nucleus' size and surface. It can, however, lead to confusion
when comparing simulations, phenomenological models and
interpretations of experiments as these may have been analysed using different definitions. 
In particular, CNT-based models generally assume that the 
nucleus has the bulk density. 
This is equivalent to taking 
a reference state based on the molar volume, or to using a 
zero-excess-volume dividing surface. A $V$-based
value of $\gsl$ is therefore desirable 
as one can then establish a direct link
between simulations and CNT models.
In addition, simulations run with
different order parameters can be compared
straightforwardly if the volume-based interfacial free energy $\gamma^V$ is also computed.

\begin{figure}
\centering
\includegraphics[width=0.95\columnwidth]{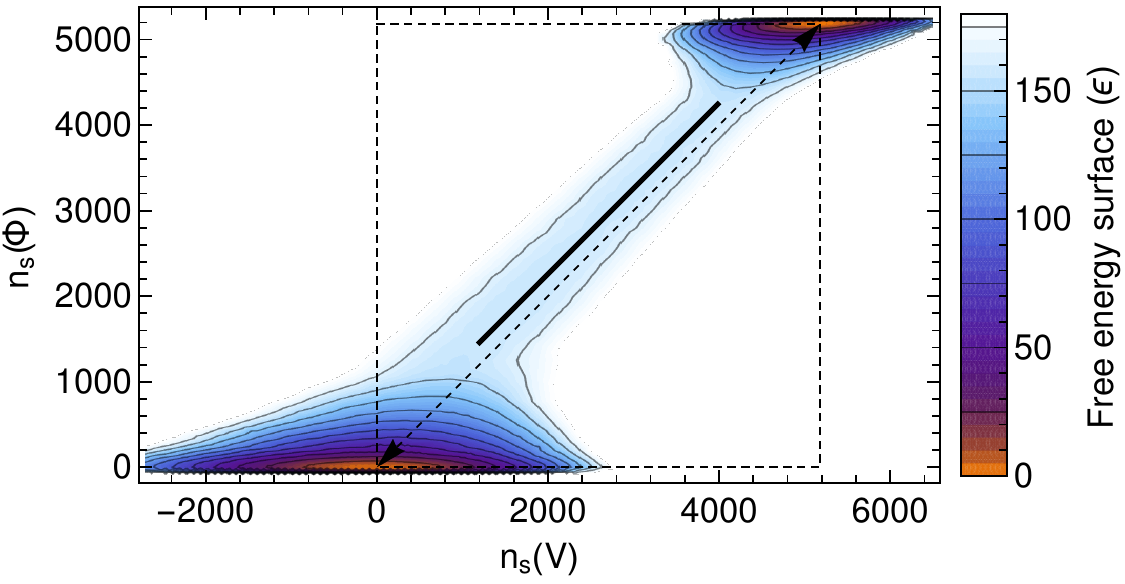}
    \caption{
    The free energy as a function of $\ns(\Phi)$ and $\ns(V)$
    for a simulation at $\tm=0.6185$ computed using a supercell
    $16\times 9 \times 9$ times the
    \emph{fcc} unit cell.
    The dashed line
    indicates the  $\ns(\Phi)=\ns(V)$ ideal line, whereas
    the full line indicates the mean value of $\ns(V)$ for a given
    value of  $\ns(\Phi)$. The horizontal offset between the two curves 
    corresponds to the surface excess for the atom count associated with 
    the $\Phi$-based dividing surface,
    $\Delta \ns\left(\Phi,V\right)$, which for this particular calculation is equal to  $259\pm 2$ atoms.
    \label{fig:fes2d}
}
\end{figure}

In principle it is possible to compute $\gsl$ using 
a different local order parameter $\theta$ 
by on-the-fly reweighing the biased trajectory. 
In practice, however, any
results obtained are only 
reliable when the sampling of  
$\Theta=\sum_i \theta(i)$ is as thorough as the sampling of
the CV used in the metadynamics simulations.
In addition, in order to extract $\gsl$, 
the free energy profile 
as a function of $\ns(\Theta)$ must
exhibit a clear plateau region that 
corresponds to solid-liquid coexistence. 
Unfortunately, the molar volume does not 
fulfill these two criteria.
To illustrate this problem, we performed 
2D metadynamics simulations in 
which $\Phi=\sum_i S(\kappa(i))$ 
(see above) and the total volume $V$ were used as CVs.  
In Figure 2 the resulting free energy surface 
is displayed as a function of the solid atom 
counts obtained when these two extensive 
quantities are inserted into Eq.~\eqref{eqn:nsl-phi}.
The fluctuations in the molar volume for the two bulk 
phases are so large that they overlap with the plateau
region. The molar volume, by itself, is thus
not an effective fingerprint for distinguishing 
the solid and the liquid phases at the atomic scale.

Even though one cannot immediately obtain $\gsl^{V}$ 
from the free energy profile as a function 
of $\ns(V)$, it can still be determined indirectly.  
To do so, one simply computes 
the change in atom count that occurs when $V$ 
rather than $\Phi$ is used to 
define the reference state:
\begin{equation}
\Delta \ns\left(\Phi,V\right) = 
\left< \ns(V) \delta\left[\ns(\Phi)-\ns(\Phi')\right]  \right> - \ns(\Phi).\label{eqn:delta-ns}
\end{equation}
In the above, the Dirac $\delta$ ensures that only states
with a certain value of $\ns(\Phi)$ are selected, and $\left< \cdot \right>$ represents a NPT ensemble average, which can be obtained by 
appropriately re-weighting the biased simulation.
$\Delta \ns(\Psi,V)$ 
can be determined to a very high statistical accuracy as long as one of the two order parameters
can identify the coexistence region. Graphically, this quantity
corresponds to the  horizontal
offset between the solid black line
and the dashed diagonal in Figure~\ref{fig:fes2d}, which is constant 
 over the entire plateau region.
This implies that, when a solid-liquid 
interface is present, 
a change in the definition of the 
Gibbs dividing surface leads to a
constant shift in $\ns$. 
The interfacial free-energy based on the volume, $\gsl^{V}(T)$,
can thus be inferred from the quantity 
$\gsl^{\Phi}(T)$ using 
$\gsl^{V}=\gsl^{\Phi}-\msl \Delta \ns\left(\Phi,V\right)/A$.

\begin{figure}
\centering
\includegraphics[width=0.5\textwidth]{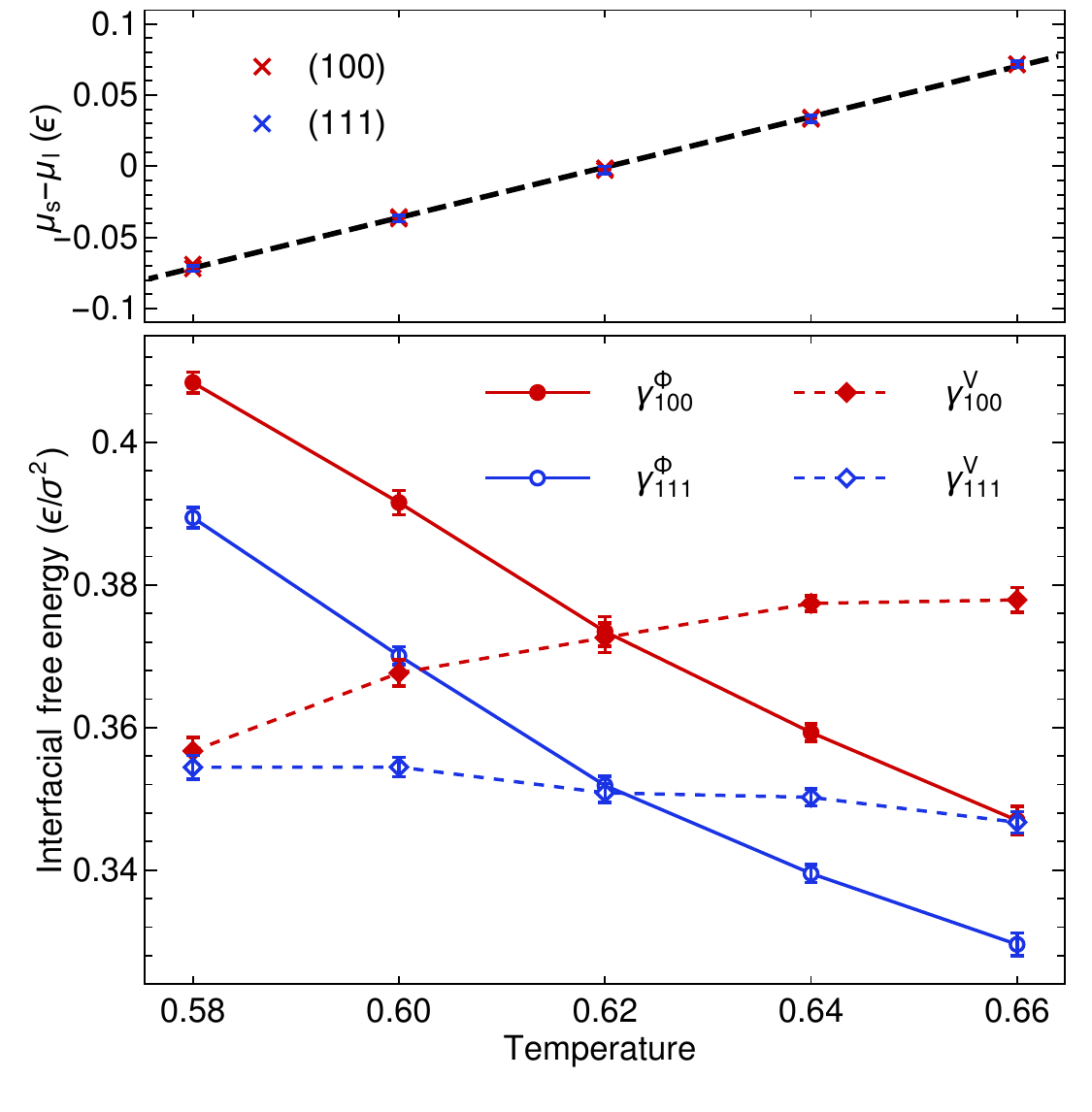}
    \caption{
   $\msl=\ms-\ml$ (upper panel) and $\gsl$
   (lower panel) for two different crystal orientations, as a function of temperature.
   $\gsl^{\Phi}$ and $\gsl^{V}$ were computed using a Gibbs dividing surface that has zero surface excess for
    $\Phi$ and $V$, respectively (see Eq.~\eqref{eqn:nsl-phi}).
    These two different ways of partitioning the free energy of the system between bulk and surface terms ensure that there are significant differences between $\gsl^{\Phi}$ and $\gsl^{V}$ when $T\ne\tm$.
 Each data point represents an average over 24 independent metadynamics runs and statistical uncertainties are indicated by the error bars.
 The uncertainties in $\gsl^{\Phi}$ and $\msl\Delta \ns\left(\Phi,V\right)$ were assumed to be independent during the error estimation.
 The data points were obtained by restricting 
the sampling to prevent complete melting, as discussed in the Supplemental Material~\cite{SI}.
 The simulation boxes comprised about 1200 atoms, 
 which implies a small finite-size effect that shifts $\tm$ to 
 $0.62$. At $\tm$ the interfacial free energies, for the two orientations we considered are $\gamma_{100}(0.62)=0.373(2)$ and $\gamma_{111}(0.62)=0.352(1)$.
All quantities are expressed in Lennard-Jones units.
   \label{fig:gslt}
   }
\end{figure}

Figure~\ref{fig:gslt} shows the temperature 
dependence for
$\msl$ and $\gsl$ computed for
the $\left<100\right>$ and $\left<111\right>$ crystal lattice orientations. 
$\msl$ is a bulk property so its value should not depend on the orientation of the surface
as is observed in the top panel.
The magnitude and anisotropy of the interface free energy at $T=\tm$ match 
the values reported in the literature~Ref.~\cite{davidchack2003direct,angi+10prb,benjamin2014crystal}. 
Its temperature 
dependence, however, depends strongly on the choice of reference state.
The lower panel shows that $\gsl^{\Phi}$ has a near-constant negative slope for both orientations 
which indicates that the excess entropy associated with the $\Phi$-based dividing surface is positive.  
The importance of using a consistent reference state when studying the solid-liquid 
interface in out-of-equilibrium conditions is
highlighted by the dramatically different 
behavior observed when the volume is used to
define the reference state.
$\gamma_{100}^V$ \emph{increases} 
with temperature as $\Delta \ns(\Phi,V)$ is large and negative.
When the molar volume is used to define the reference state
a larger fraction of the interface
region is seen as an ordered liquid rather than a
disordered solid. The interface for the $\left<100\right>$
crystal orientation thus has a negative
excess surface entropy.
For the $\left<111\right>$ 
orientation, meanwhile, $\Delta \ns(\Phi,V)$
is smaller and the slope of $\gamma^{V}_{111}$ 
remains negative.

It is important to remember that the total 
free energy of the nucleus is the only quantity 
that affects its thermodynamic stability.
However, the dependence of $\gsl$ on the dividing surface will have consequences 
on predictions and data analysis if the same reference 
state is not used consistently. 
Usually, CNT models are built assuming that the 
nucleus' density is constant and equal to
its bulk value. To be consistent with this
assumption, one should thus use $\gsl^V$ 
when making predictions using this theory. 
In the Supplemental Material at [URL will be inserted by publisher]~\cite{SI} 
we show that, when using a reference state with a non-zero surface 
volume excess, additional correction terms must be included on top 
of the usual CNT expressions.  In addition to changing the value of  $\gsl$,
one must use a density that differs from the bulk value when inferring the surface area of the cluster. This density depends on the nucleus size in a manner 
that is reminiscent of the well-known Tolman correction, which has been 
shown to be effective when extrapolating the value of $G^\star$ obtained 
from simulations of nanoscopic nuclei to the mesoscale~\cite{prestipino2012systematic}.   

The  observation of a positive 
slope for $\gsl^V(T)$ is consistent with 
a simple phenomenological model 
for a hard-sphere system
~\cite{spaepen1994homogeneous},
and with the effective $\gsl$ values
that were determined by fitting 
the values of $G^\star(T)$  obtained from simulations
run at deeply undercooled conditions using a CNT expression~\cite{trudu2006freezing}.
Our results paint a somewhat more nuanced picture, however.
The behavior of $\gamma^{V}_{100}$ deviates significantly from linearity even at small undercoolings.  In addition,
the temperature dependence of the 
solid-liquid interface energy computed relative 
to a zero-excess-volume reference
depends strongly on orientation, with the
$\left<111\right>$  direction showing a small
negative slope.
As a consequence, the anisotropy in $\gsl^V$ 
between the two orientations we considered decreases down 
to $T=0.58$, suggesting that the shape of the critical nucleus 
may depend on the degree of undercooling. 
Fully determining the size and shape of the 
critical nucleus would require one to determine
$\gsl$ and $\partial\gsl/\partial T$ for other
high-symmetry orientations and for small 
mis-orientations. A direct comparison between 
planar-interface calculations and 3D nucleation
simulations at intermediate undercoolings would also allow one to 
test the validity of the various assumptions within
CNT.   For example, such simulations would allow one to determine whether or not
the Laplace pressure caused by the curvature of the interface has a significant effect on the chemical potentials of the two phases.
These calculations, as well as the investigation of more realistic 
inter-atomic potentials, will be the subject of future work, which will provide
more accurate parameters for
mesoscale phase-field models of nucleation 
and growth~\cite{curtin1987density,chen2002phase}. 

Using a thermodynamic definition
for the size of the solid region 
allows one to treat atomic-scale fingerprints
and macroscopic observables on the same footing, 
thereby providing a practical 
procedure for converting the value of 
$\gsl$ obtained with a 
computationally-effective order parameter 
into the value consistent with the assumptions 
of mesoscopic theories of nucleation.
The combination of this framework and 
accelerated molecular dynamics makes 
it possible to generate and stabilize 
planar interfaces at $T \ne \tm$.  
Consequently, simulations can be 
performed with conditions that more 
closely resemble those in experiments.
Such simulations could be used to shed 
light on the changes 
in morphology of the critical nucleus, to 
monitor capillary fluctuations at $T\ne\tm$ 
\cite{hoyt2001method}, and to give atomistic
insight on the evolution of interfaces away
from equilibrium\cite{gurt-voor96am}.
In addition, the methodology and the thermodynamic 
considerations we make here,
whilst examining solidification,
would apply with minor modifications to the study
of nucleation in other contexts, 
be it melting~\cite{sama+14science},
precipitation~\cite{salv+12jacs},
condensation,  order-disorder transitions, 
or other situations in which a phase 
transition is hindered 
by a surface energy term\cite{khal+11nmat}. 

We would like to thank Michel Rappaz for insightful discussion.
MC acknowledges the Competence
Centre for Materials Science and Technology (CCMX) for funding.

\bibliography{biblio,others} 

\end{document}